\documentclass[12pt, draftclsnofoot,onecolumn]{IEEEtran} 

\usepackage{amsmath}
\usepackage{amssymb}
\usepackage{amsfonts}
\usepackage{subfigure}
\usepackage{float}
\usepackage{cite}
\usepackage{color}
\usepackage{hyperref}
\usepackage{graphicx}
\usepackage{float}
\usepackage{algorithm}
\usepackage{algorithmic}

\begin{document}

\title{Optimized Frameless ALOHA for Cooperative Base Stations with Overlapped Coverage Areas}

\author{\IEEEauthorblockN{Shun Ogata (Corresponding Author),~\IEEEmembership{Student Member,~IEEE,}\\ Koji Ishibashi,~\IEEEmembership{Member,~IEEE,} and Giuseppe Abreu,~\IEEEmembership{Senior Member,~IEEE}}
  \thanks{Part of this material appeared at the 49th Asilomar Conf. on Sigs., Syst., and Comput. (see \cite{Asilomar}), and the 2017 IEEE 18th Int. Workshop on Signal Process. Advances in Wireless Commun. (see \cite{SPAWC}).}
  \thanks{S. Ogata and K. Ishibashi are with the Advanced Wireless \& Communication Research Center (AWCC), University of Electro-Communications, 1-5-1 Chofugaoka, Chofu-shi, Tokyo 182-8585, Japan (e-mail: shun@awcc.uec.ac.jp, koji@ieee.org)}
  \thanks{G. Abreu is with the School of Engineering and Sciences, Jacobs University Bremen, Campus Ring 1, 28759 Bremen, Germany and Department of Electrical and Electronic Engineering, College of Science and Engineering, Ritsumeikan University, 1-1-1 Nojihigashi, Kusatsu-shi, Siga 525-0058, Japan (e-mail: g.abreu@jacobs-university.de; g-abreu@fc.ritsumei.ac.jp)}%
}

\maketitle

\vspace{-6ex}
\begin{abstract}
Herein, we consider the problem of cooperative multi-access in the presence of overlapped coverage areas.
Assuming a frameless ALOHA transmission scheme, we derive exact analytical throughput expressions for throughput in the aforementioned scenarios as a function of the frame length of the system and for arbitrary average numbers of users transmitting in each slot (target degree).
After obtaining these original expressions, we then formulate a utility function whose maximization (obtained, $e.g.$, through genetic algorithms) yields unequal and optimum target degrees to be employed by users in each group in order to maximize the peak throughput of the whole system, while satisfying a given prescribed outage.
A comparison of the resulting cooperative multiple base station (BS) multi-access scheme against optimized single-BS frameless ALOHA systems --- which presume the perfect isolation of users at each BS and an equal optimum target degree for all users --- indicates a significant gain in overall throughput, thereby revealing that a ``\emph{multi-access diversity gain}'' can be reaped by allowing groups of users from different BSs to overlap.
\end{abstract}

\vspace{-2ex}
\begin{IEEEkeywords}
  Frameless ALOHA, successive interference cancellation, multiple base station cooperation
\end{IEEEkeywords}
\vspace{-2ex}

  \section{Introduction}\label{intro}
Providing network access to massive numbers of devices is a capability that is gathering interest as the Internet-of-Things (IoT) continues to be developed and approaches deployment.
In such networks, random access approaches are more effective than resource-assignment access schemes such as {\em time division multiple access} (TDMA), since the controlled assignment of channel resources to large numbers of users leads to prohibitive overheads.
Slotted ALOHA systems employing successive interference cancellation (SIC), also known as {\em coded ALOHA}, have been well studied and have been shown to achieve excellent throughput performance comparable to TDMA systems \cite{CRDSA,Liva,Peter_lett}.
On the other hand, most coded ALOHA studies only consider the use of a single base station (BS), which goes against the recent trend of introducing multiple BSs into large wireless networks in order to increase their spectrum efficiency\cite{Shin}.

Recently, multi-access schemes using multiple BSs have been proposed as a means to increase throughput.
In \cite{Qian}, an interference cancellation-based non-orthogonal multiple-access (NOMA) scheme with multiple BSs was proposed. That scheme exploits the difference between received power among multiple BSs in order to retrieve colliding packets, and jointly optimizes the power allocation of users as well as the BS association to ensure that the capacity of each BS is maximized while minimizing total transmission power.
In addition, an opportunistic transmission method in a multi-cell network is discussed in \cite{Lin17ISIT}, where each user simultaneously considers whether 1) the channel gain to the receiver is sufficiently large, and 2) the interference that a user causes to other receivers is sufficiently small.
These schemes attempt to maximize network throughput by considering the signal power and signal-to-interference-plus-noise ratio (SINR), which is achieved by focusing on the physical layer rather than the medium access control (MAC) layer itself.
When it comes to the MAC layer, inter-slot SIC (hereafter termed SIC) over a multiple-BS network was recently studied by \cite{Jakovetic}.
In that study, the authors proposed improving SIC performance via multiple BS cooperation via a system in which users transmit their packets using the {\em framed} ALOHA strategy described by \cite{Liva}, and BSs share retrieved packets using a backhaul network that allows each BS to cancel the shared packets.
Moreover, supposing that users and BSs are deployed following a Poisson point process (PPP), analytical expressions for the packet loss rate (PLR) and throughput are derived, with the analysis results used to optimize the number of user retransmissions.
The main conclusion reached in \cite{Jakovetic} is that it is nearly optimal for all the users to transmit twice during the frame.

However, while the work of \cite{Jakovetic} is very informative, it faces several practical limitations.
For instance, regarding the user transmission protocol, the framed structure requires a suitable number of time slots, which is a challenging requirement for multiple-BS networks since each BS will have a different number of users.
Moreover, the optimized transmission strategy might be sub-optimal in practice due to the PPP assumption.
Furthermore, the proposed optimization method presupposes that the average number of BSs connected to each user is identical, while in practice the number of BSs and the retransmission numbers are different for each user.

In this paper, we derive an exact theoretical PLR expression for coded ALOHA with multiple BS cooperation.
Specifically, we assume the use of frameless ALOHA \cite{Peter_lett}, which is a recently proposed coded ALOHA scheme in which the frame length is automatically determined on the fly under the constraint that sufficiently large packet numbers are retrieved.
This frameless ALOHA scheme is well suited for multiple-BS networks because its {\em frameless} structure is realized via probabilistic user retransmissions that are based on a given transmission probability, thus avoiding the problem of frame length determination that arises in framed schemes.
In contrast to the approximated analysis of \cite{Jakovetic}, the exact analysis takes into account the connectivity of users and BSs so that packet sharing among BSs can be precisely tracked.
Our proposed analysis is then employed to optimize the average number of user retransmissions so as to maximize the network throughput.
A comparison of the resulting cooperative multiple-BS multi-access scheme against optimized single-BS frameless ALOHA systems -- which presumes perfect user isolation at each BS and an equal and optimum average number of retransmissions for all users -- indicates a significant gain in the overall throughput, thus revealing that a ``\emph{multi-access diversity gain}'' can be reaped by allowing user groups from different BSs to overlap.
The theoretical analysis of throughput is further used to quantitatively evaluate the multi-access diversity gain, with the results confirming that the gain grows as the number of BSs increases.
Numerical examples demonstrate that the proposed frameless ALOHA with optimized parameters exhibits higher maximum throughput performance than the state-of-the-art multiple BS random access scheme proposed in \cite{Jakovetic}.
The contributions of this paper are summarized as follows:
\begin{itemize}
\item We derive an exact theoretical throughput expression for frameless ALOHA with multiple BS cooperation and demonstrate its application to the optimization of transmission probabilities in order to maximize the achievable throughput.
\item Simple lower and upper bounds for cooperative throughput are introduced to calculate the throughput performance for a large number of BSs in order to guarantee that the multi-access diversity gain increases as the number of BSs increases.
\item The results show that our proposed scheme outperforms the conventional scheme in \cite{Jakovetic} in terms of maximum throughput.
\end{itemize}

The remainder of the paper is organized as follows.
Section \ref{model} introduces the system model and describes frameless ALOHA with multiple BS cooperation.
In section \ref{analysis}, a theoretical throughput analysis of frameless ALOHA with and without multiple BS cooperation is presented. The results are then used to obtain the multi-access diversity gain.
While the analysis results can be applied to an arbitrary number of BSs, we specifically demonstrate its utility in a three-BS network.
Moreover, the upper and lower bounds of the multi-access diversity gain are introduced.
Section \ref{numerical_example} shows some numerical results related to the target degree optimization and the average throughput performance, as well as a comparison with a state-of-the-art scheme named {\em spatio-temporal cooperation} \cite{Jakovetic}. Those results show that our proposed scheme achieves a higher throughput performance.
Finally, we conclude the paper in Section \ref{conclusion}.
\section{System Model}\label{model}
\subsection{Notations}
Calligraphic letters will be used to denote sets, and we will denote the set differences between $\mathcal{S}_1$ and $\mathcal{S}_2$ by $\mathcal{S}_1\setminus\mathcal{S}_2$.
A san-serif font is used to indicate user group indexes and BSs. For instance, $\mathsf{u}_i$ denotes the $i$-th user group, and $\mathsf{s}_j$ corresponds to the $j$-th BS.
Let $\lfloor\cdot\rfloor$ denote a floor function and $f'(x)$ be the first derivative of the function $f$ with respect to $x$. Vectors and matrices are denoted by bold-faced characters, and the probability of a random event $A$ is denoted by $\mathrm{Pr}(A)$.
\subsection{Network Model}
\begin{figure}[t]
  \centering
  \includegraphics[width=0.4\hsize]{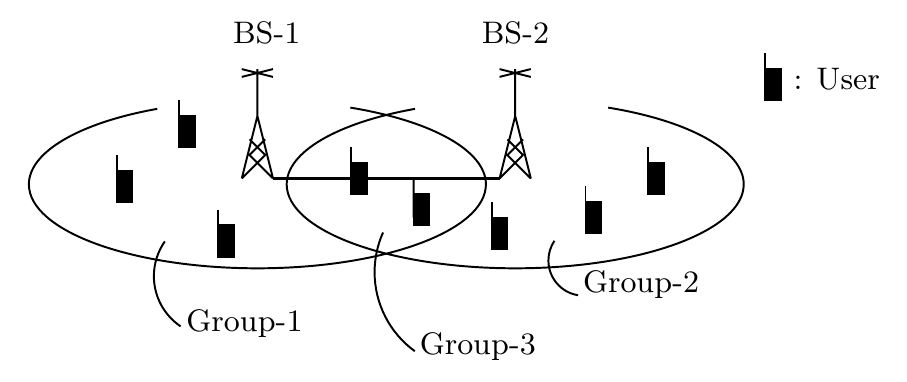}
  \vspace{-4ex}
  \caption{An example of a network model for $M=2$. Users in group-3 $(\mathsf{u}_3)$ can communicate with both BSs, which are connected via a backhaul network.}
  \label{model_2BSs}
  \vspace{-2ex}
\end{figure}

Throughout this paper, we will consider a network with $N$ users and $M$ BSs.
Each user has one packet at the start of a frame, and no new packets are generated during the frame.
Moreover, users do not have specific destinations and strive to deliver their packets to any BSs that can receive the packets.
Each user transmits the same packet in all the time slots of the frame.
Users are categorized into multiple groups depending on which BS(s) they are able to communicate with.
Let $I$ denote the number of user groups.
It is assumed that each user can communicate with at least one BS.
Thus, there exists at most $2^M-1$ user groups, where $N_i$ denotes the number of users in the $i$-th user group.
Hereinafter, $\mathsf{u}_i$ denotes the $i$-th user group, and $\mathsf{s}_j$ denotes BS-$j$.
Figure \ref{model_2BSs} shows an example of a network model for $M=2$.

Not only do users need to be associated with the network, they also need to know which user group they are participating in and the number of users in the group in order to calculate the transmission probability, as will be described later.
To this end, each user at first broadcasts a short packet to the BSs in order to announce its intention to participate in the network prior to transmitting data.
This short packet transmission is conducted before the data transmission, and we assume that the network association will be ideally finished, $i.e.$, the short packet is received at the BSs without any errors.
The BSs share the received short packets with each other in order to calculate the number of users for each group.
Finally, the BSs report the number of users in each group, and which group users belong, to the users.
Although users may be allocated randomly in practice, this paper only considers an analysis for a deterministic allocation because our analysis method can be applied to any network simply by changing the number of users in each group.
It is assumed that all of the BSs are connected to each other via a backhaul network, so they can communicate with each other without errors.
Additionally, all the users and BSs are assumed to be temporally synchronized so that each transmission occurs in a time slot.

\subsection{Frameless ALOHA Transmission}
In every time slot, each user decides whether or not to transmit its own packet using a transmission probability.
The transmission probability of $\mathsf{u}_i$ is given by $p_i=G_i/N_i$, where $G_i$ is called the {\it target degree}, which is defined as the average number of users in $\mathsf{u}_i$ transmitting in one time slot.
As stated above, before they can calculate $p_i$, users need to know their group index $i$ and $N_i$.
Target degrees can be arbitrarily set and are shared among each user group.
Although target degrees can be changed at every time slot, as mentioned in \cite{Peter_lett}, that study also shows that even a constant target degree yields a throughput performance comparable with that for multiple target degrees.
Thus, hereafter we only consider constant target degrees for each user group.

\begin{figure}[t]
\centering
\includegraphics[width=\hsize]{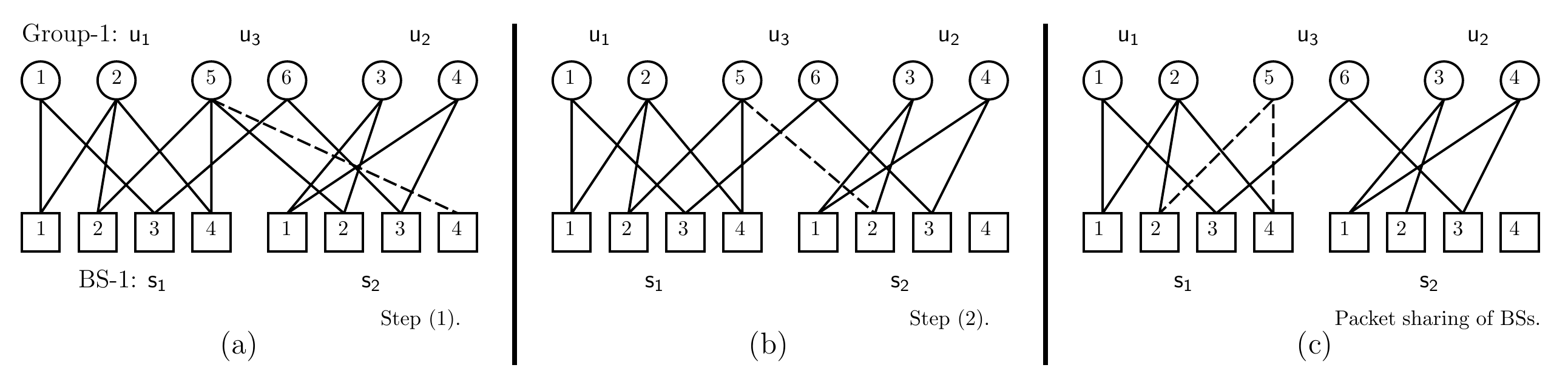}
\vspace{-6ex}
\caption{Transmission graph and SIC. (a) An example of a transmission graph. The user-5 packet can be retrieved at $\mathsf{s}_2$. (b) The user-5 packet is subtracted from corresponding received packets. (c) The use of multiple BS cooperation makes it possible for $\mathsf{s}_1$ to subtract the user-5 packet from the corresponding received packets.}
  \label{graph_2BSs_sic}
  \vspace{-2ex}
\end{figure}

Frameless ALOHA transmission can be represented by a bipartite graph that consists of edges and two kinds of nodes: {\it variable nodes} and {\it observation nodes}.
Variable nodes and observation nodes correspond to information packets and received packets, respectively.
Note that since $M$ BSs in the network simultaneously receive packets, $M$ observation nodes exist for each time slot.
When the frame length is $T$, there are $MT$ observation nodes in the transmission graph.
The user transmission is depicted by an edge between the variable and observation nodes.
The number of edges connected to each node is called the {\it degree} of the node. Note that the degree of a variable node corresponds to the number of user transmissions in the frame, while the degree of an observation node corresponds to the number of users transmitting in the time slot.
Hereinafter, we refer to the bipartite graph as a {\em transmission graph} in order to emphasize that the graph visually represents the user {\em transmissions}.
In order to help readers to visually understand the bipartite graph, an example of a transmission graph with $M=2$ BSs is shown in Fig. \ref{graph_2BSs_sic}.
In the example, user-5 belonging to $\mathsf{u}_3$ transmits the packet in slot-2 and slot-4.
Since packets transmitted by users in $\mathsf{u}_3$ can be received at both $\mathsf{s}_1$ and $\mathsf{s}_2$, the variable node of user-5 is connected with observation nodes of slot-2 and slot-4 of $\mathsf{s}_1$ and $\mathsf{s}_2$.

\subsection{Packet Retrieval with Multiple Base Station Cooperation}
Since users transmit their packets independently, some users may transmit simultaneously in the same slot, thus leading to {\em packet collisions}.
Packets that collide are considered to be lost since, for mathematical tractability reasons, the capture effect \cite{capture} is not considered.
Since this model is considered to be the worst-case scenario, it provides a lower-bound to throughput performance in practical situations where the capture effect would be available. 
Hence, the model is both relevant and useful  because it permits an achievable performance to be obtained, and has been used in numerous conventional studies, $e.g.$ \cite{CRDSA,Liva,Jakovetic}.
It is further assumed that BSs are able to distinguish the conditions for each time slot:
(i) no users have transmitted,
(ii) only one user has transmitted, $i.e.$, the time slot is a singleton, and
(iii) some users have transmitted and packets have collided.
However, even if the BS detects a collision, it cannot identify which user's packets have collided, or how many packets have collided.
Frameless ALOHA employs SIC in order to retrieve the original packets from collisions.
The SIC for frameless ALOHA is equivalent to the peeling decoder for low-density parity check (LDPC) codes \cite{Urbanke}.
As depicted in Fig. \ref{graph_2BSs_sic}, the SIC process can be described using the following steps:
\begin{itemize}
\item[(i)] Retrieve the transmitted packets from singleton slots as depicted in Fig. \ref{graph_2BSs_sic}--(a). The slots are assumed to be empty.
\item[(ii)] Subtract the packets from all the received signals in which the packets are included.
\end{itemize}
After step (ii), some collided packets become singletons, $e.g.$, the second time slot of $\mathsf{s}_2$ becomes a singleton in Fig. \ref{graph_2BSs_sic}--(b), and the above operations are repeated until all the singleton slots vanish.
In order to execute step (ii), it is assumed that each packet includes information indicating the time slot in which it is transmitted.
Note that the retrieved packets might be included in future received packets.
To this end, if each user identification (ID) is used as a seed for a random generator for choosing time slots in which to transmit, the receiver can determine all the future transmissions and subtract signals from all the received packets \cite{Ricciato}.
We assume that a backhaul network is used among the participating BSs in order to immediately share successfully retrieved packets, so that the received packets are subtracted from all the received signals in which shared packets are included, thus leading to additional singleton time slots, as shown in Fig. \ref{graph_2BSs_sic}--(c).
The combination of shared packets among BSs with SIC is equivalent to the decoding process in {\it spatio-temporal cooperation}, which was proposed by \cite{Jakovetic}.
In frameless ALOHA, the frame length can be arbitrarily extended to retrieve more transmitted packets.
However, a large number of time slots may be required to retrieve all the transmitted packets, which could result in significant delays.
In this paper, it is assumed that the frame is terminated when $\lfloor\alpha N\rfloor$ packets are successfully retrieved, thus providing a point where the threshold $\alpha\in (0,1]$ can be set arbitrarily.

Upon transmission, slots are organized into a {\it{frame}}, with the start and end of the frame reported by the receiver via a beacon.
The name {\it frameless} comes from the fact that the frame length is not {\it a priori} fixed.
Users whose packets were not retrieved in the previous frame will retransmit the same packets in the next frame.
Thus, there is only one feedback from BSs, where the feedback signal stops user transmissions.
To this end, after each slot, it is presumed that the users will wait a short period for the feedback signal.
If users do not receive the feedback signal, they continue the frameless ALOHA transmission.
\subsection{Degree Distributions}
Let us denote the number of time slots by $T$.
{\em Degree distributions} characterize the randomly constructed transmission graph and can be used to theoretically analyze the PLR for frameless ALOHA.
We define $L_{i,k}$ as the probability that the variable node for a user in $\mathsf{u}_i$ has a degree-$k$, $i.e.$, that the user in $\mathsf{u}_i$ has transmitted the packet $k$ times during $T$ slots.
The probability $L_{i,k}$ is given by
\begin{equation}
  L_{i,k}=\binom{T}{k}p_i^k(1-p_i)^{T-k}.
\end{equation}

{\color{black}Furthermore, $R_{i,k}$ is the probability that $k$ users in $\mathsf{u}_i$ transmit in the slot.}
Then, the probability is given by
\begin{equation}
  R_{i,k}=\binom{N_i}{k}p_i^k(1-p_i)^{N_i-k}.
\end{equation}

Using the probabilities, {\em node-perspective} degree distributions are defined as
\begin{equation}
  L_i(x)\triangleq \sum_{k=0}^TL_{i,k}x^k,
\end{equation}
and
\begin{equation}
  R_i(x)\triangleq \sum_{k=0}^{N_i}R_{i,k}x^k,
\end{equation}
where $x$ is a dummy variable.

The node-perspective degree distributions yield {\em edge-perspective} degree distributions as
\begin{equation}
  \lambda_i(x)\triangleq\sum_{k=1}^T\lambda_{i,k}x^{k-1}=L_i'(x)/L_i'(1),\label{eq:lambda}
\end{equation}
and
\begin{equation}
  \rho_i(x)\triangleq\sum_{k=1}^{N_i}\rho_{i,k}x^{k-1}=R_i'(x)/R_i'(1).\label{eq:rho}
\end{equation}

Hereafter, for the sake of brevity, we will use the symbol $\mathsf{u}_i$ to refer either the user group or a packet of the user in the group, according to context.
{\color{black}
The coefficient in \eqref{eq:lambda}, namely $\lambda_{i,k}$, denotes the probability that the transmitted packet $\mathsf{u}_i$ is retransmitted $k$ times during the frame.
Similarly, $\rho_{i,k}$ in \eqref{eq:rho} denotes the probability that the transmitted packet $\mathsf{u}_i$ has collided with other $(k-1)$ packets transmitted by users in $\mathsf{u}_i$ in the slot.
}


\section{Throughput Analysis}\label{analysis}
In this section, we will attempt to quantitatively determine how much performance improvement can be achieved via multiple BS cooperation.
To this end, we derive theoretical expressions for the frameless ALOHA throughput with and without multiple BS cooperation.
Given a number of time slots $T$, the throughput $S(T)$ is defined as the fraction of successfully retrieved packets and time slots, and is given by
\begin{equation}
  S(T)\triangleq \frac{N_{\rm ret}(T)}{T},\label{definition_throughput}
\end{equation}
where $N_{\rm ret}(T)$ denotes the number of retrieved packets within $T$ slots, and $0\leq N_{\rm ret}(T)\leq N$.

Moreover, we introduce a metric named {\em multi-access diversity gain} to evaluate the performance improvement achieved via multiple BS cooperation.
We define $S^{\rm c}$ and $S^{\rm nc}$ as the throughput performance for frameless ALOHA with and without multiple BS cooperation, respectively.
Then, the multi-access diversity gain $\Gamma$ is defined as
\begin{equation}
  \Gamma\triangleq \frac{S^{\rm c}}{S^{\rm nc}}.\label{eq:gain}
\end{equation}

To theoretically calculate throughput performance, $N_{\rm ret}(T)$ should be theoretically obtained in \eqref{definition_throughput}.
To this end, denoting by $p_{\rm e}(T)$ the PLR with $T$ time slots, $N_{\rm ret}(T)$ is given by $N_{\rm ret}=N(1-p_{\rm e}(T))$, where the PLR $p_{\rm e}(T)$ needs to be theoretically derived.
In the following subsection, a theoretical PLR expression for frameless ALOHA with multiple BSs is derived while considering two scenarios: non-cooperative BSs and cooperative BSs.

\subsection{Analysis of Non-Cooperative Packet Retrieval}
In situations without multiple BS cooperation, each BS locally attempts to retrieve transmitted packets.
Since packets are retrieved via an iterative SIC process, a theoretical PLR can be obtained via iterative calculations.
{\color{black}
  The idea behind the calculation is similar to the original density evolution \cite{Urbanke}.
Specifically, in order to obtain the PLR of $\mathsf{u}_i$, two kinds of variables, namely $x_{i,j}^{(l)}$ and $w_{i,j}^{(l)}$, are iteratively calculated for each $j$ in a way that ensures the users in $\mathsf{u}_i$ can communicate with $\mathsf{s}_j$.
  The former, $x_{i,j}^{(l)}\in[0,1]$, is the probability that the packet $\mathsf{u}_i$ will not be retrieved at $\mathsf{s}_j$ in the $l$-th iteration. This event occurs when all the retransmitted packets have collided.
  Then, $x_{i,j}^{(l)}$ is given by
  
  \begin{equation}
    x_{i,j}^{(l)}=\left\{
    \begin{array}{l}
      \lambda_i(w_{i,j}^{(l)}),\label{x_ij}\ \ {\rm for}\ l\geq 1\\
      1,\ \ {\rm for}\ l=0,
    \end{array}
    \right.
  \end{equation}
  where $x_{i,j}^{(0)}=1$ indicates that no packet has been retrieved at the beginning of the retrieval process, and $\lambda_i(w_{i,j}^{(l)})=\sum_{k=1}^{T}\lambda_{i,k}\times (w_{i,j}^{(l)})^{k-1}$ corresponds to the probability that all the incoming $(k-1)$ edges are still colliding.
  
The latter, $w_{i,j}^{(l)}\in[0,1]$, is the probability that the packet $\mathsf{u}_i$ has collided at $\mathsf{s}_j$ in the $l$-th iteration.
To calculate $w_{i,j}^{(l)}$, we consider the probability that the transmitted packet becomes un-collided.
The probability of all users in $\mathsf{u}_i$ except the specified user being retrieved is $\rho_i(1-x_{i,j}^{(l-1)})=\sum_{k=1}^{N_i}\rho_{i,k}\times(1-x_{i,j}^{(l-1)})^{k-1}$, and the probability that all users in $\mathsf{u}_m$ have been retrieved is $R_m(1-x_{m,j}^{(l-1)})=\sum_{k=0}^{N_m}R_{m,k}\times(1-x_{m,j}^{(l-1)})^k$.
  Then, $w_{i,j}^{(l)}$ is given by
  \begin{align}
    w_{i,j}^{(l)}=1-\rho_i(1-x_{i,j}^{(l-1)})\prod_{m:\mathsf{u}_m\in\mathcal{U}(\mathsf{s}_j)\setminus\{\mathsf{u}_i\}}R_m(1-x_{m,j}^{(l-1)}),\label{w_ij}
  \end{align}
  where $\mathcal{U}(\mathsf{s}_j)$ is a set of user groups established in a way that permits users in the group to communicate with $\mathsf{s}_j$.
}
After a sufficiently large number of iterations, retrieval of the packet $\mathsf{u}_i$ only fails when the packet retrieval attempt fails at all the BSs.
Hence, the theoretical PLR of $\mathsf{u}_i$, say $p_{\mathrm{e},i}(T)$, is obtained by
\begin{align}
  p_{\mathrm{e},i}(T)&=L_i(w_i),\label{pe_nc}\\
  w_i&\approx\prod_{j:\mathsf{s}_j\in \mathcal{S}(\mathsf{u}_i)}w_{i,j}^{(l)},\label{w_i_nc}
\end{align}
where $\mathcal{S}(\mathsf{u}_i)$ is a set of BSs to which the users in $\mathsf{u}_i$ are connected.

The calculation of $p_{\mathrm{e},i}(T)$ in \eqref{pe_nc} is identical to the original frameless ALOHA analysis, while the calculation of $w_i$ is approximated.
The approximation comes from the assumption in \eqref{w_i_nc}, which states that the potential of a packet $\mathsf{u}_i$ to become a singleton at each slot is independent of each BS. However, this is not actually the case since the packets of groups in the overlapped areas should be received simultaneously at multiple BSs.
Nevertheless, the use of this approximation allows us to simplify the theoretical expression, as shown in \eqref{w_i_nc}, while still providing an accurate result.

\subsection{Analysis of Cooperative Packet Retrieval}
In order to clarify how to theoretically analyze the PLR for a frameless ALOHA with multiple cooperating BSs, we derive the exact theoretical expression for the PLR for a specific case with $M=3$.
When making comparisons with the non-cooperative case, it is necessary to consider the effect of packet sharing among BSs. Thus, important additional terms appear in the analytical equation.
When $M=3$, there exists at most $2^3-1=7$ user groups.

When packet sharing is employed, a packet transmitted from the overlapped coverage area should be retrieved simultaneously at all participating BSs.
This implies that the retrieval process for each BS is {\em not actually independent}, unlike the non-cooperative case where the retrieval process is performed {\em independently} at each BS.

\begin{figure}[t]
\centering
\includegraphics[width=0.4\hsize]{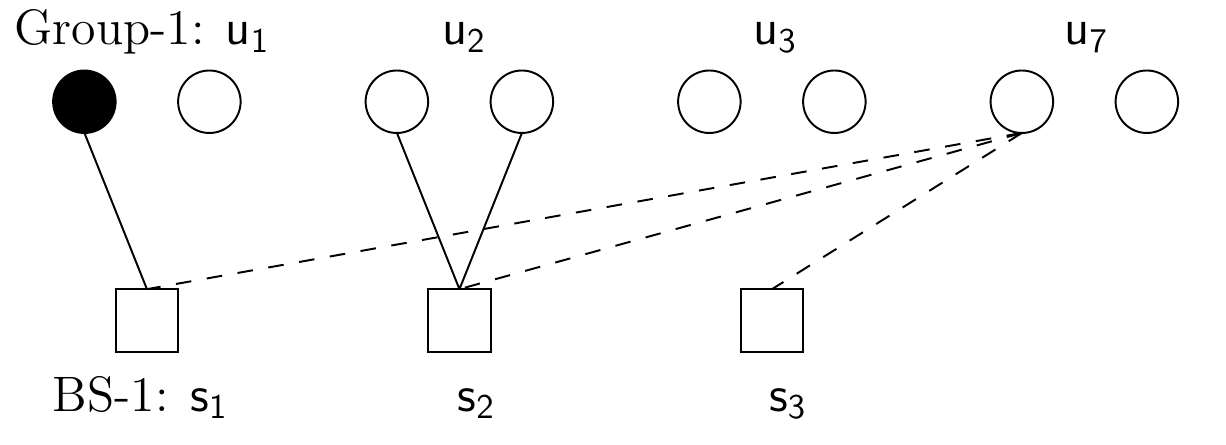}
\vspace{-4ex}
\caption{An example of a transmission graph where a collided packet $\mathsf{u}_1$ (represented by a filled circle) can be retrieved.}
  \label{edge_remain}
\vspace{-2ex}
\end{figure}
Moreover, a packet can sometimes be retrieved even if it has collided with other packets.
In Fig. \ref{edge_remain}, some users in $\mathsf{u}_1$, $\mathsf{u}_2$, and $\mathsf{u}_7$ have transmitted their packets, while presuming that no users in $\mathsf{u}_4$, $\mathsf{u}_5$, and $\mathsf{u}_6$ have transmitted packets, nor of $\mathsf{u}_3$.
Although the packet $\mathsf{u}_1$ has collided with the packet $\mathsf{u}_7$, the colliding packet is retrieved at $\mathsf{s}_3$, ultimately making it possible to retrieve the packet $\mathsf{u}_1$.

The theoretical PLR for frameless ALOHA with multiple BS cooperation obtained via iterative calculation of two kinds of variables, namely $x_i^{(l)}$ and $w_i^{(l)}$, is needed, as is the analysis of non-cooperative cases.
The calculation of $x_i^{(l)}$ is identical to the non-cooperative case, that is
\begin{equation}
  x_{i}^{(l)}=\left\{
  \begin{array}{l}
    \lambda_i(w_{i}^{(l)}),\ \ {\rm for}\ l\geq 1\\
    1,\ \ {\rm for}\ l=0.
  \end{array}
  \right.\label{x_i}
\end{equation}

As shown in Fig. \ref{edge_remain}, other important terms result when it comes to $w_i^{(l)}$with the multiple BS cooperation because of further packet retrieval.
We can describe $w_i^{(l)}$ by following the intuitive representation
\begin{equation}
  w_i^{(l)}=1-(P_i^{\rm (r0)}+P_i^{\rm (r1)}),\label{w_i}
\end{equation}
where $P^{\rm (r0)}_i$ is the probability that the packet is free from collision (also included in the analysis of non-cooperative case), and $P^{\rm (r1)}_i$ is an additional term, $i.e.$, the probability that the packet can be retrieved after it has collided with other packets.

The variable $w_i^{(l)}$ can be obtained by finding all the patterns where the packet $\mathsf{u}_i$ has been retrieved.
Specific representation of $w_i^{(l)}$ for three-BS cooperation appears in the Appendix, and an algorithm that can be used to calculate $w_i^{(l)}$ for an arbitrary number of BSs will be discussed in the next section.
For a sufficiently large $l$, the PLR for $\mathsf{u}_i$ is theoretically calculated as
\begin{equation}
  p_{{\rm e},i}(T)=L_i(w_i^{(l)}),\label{eq:plr}
\end{equation}
and the average PLR is given by
\begin{equation}
  p_{\rm e}(T)=\sum_{i=1}^I\frac{N_i}{N}p_{{\rm e},i}(T). \label{eq:avg_plr}
\end{equation}

\subsubsection{Generalized Analysis}
The equations \eqref{x_i}, \eqref{w_i}, \eqref{eq:plr}, and \eqref{eq:avg_plr} still hold for general cases of $M>1$.
Moreover, we can straightforwardly calculate \eqref{x_i}, \eqref{eq:plr}, and \eqref{eq:avg_plr} for any given network with an arbitrary $M$.
The calculation of \eqref{w_i}, however, cannot be solved in a straightforward manner, as $P_i^{\rm (r0)}$ and $P_i^{\rm (r1)}$ consist of a large number of probabilities indicating specific patterns where the specified packet can be retrieved.
The specific form of $w_i^{(l)}$ depends on the number of BSs and the consequent network topology, and can only be obtained by searching all the cases where the packet $\mathsf{u}_i$ can be retrieved.
Since the number of terms included in $P_i^{\rm (r0)}$ and $P_i^{\rm (r1)}$ is large\footnote{For instance, when $M=4$, the number of terms included in $P_i^{\rm (r0)}$ is 69,356, and the number of terms included in $P_i^{\rm (r1)}$ is 6,108\cite{SPAWC}.}, it is important to take all the patterns into consideration in order to fully reveal the throughput performance.
If some patterns are ignored in the analysis, the analytical performance becomes worse than the practical performance.
To this end, we introduce a {\em walk graph} representation of multiple BS networks that visualizes a snapshot at the time slot.
The walk graph consists of two kinds of nodes, namely user nodes and BS nodes, which correspond to user groups and BSs, respectively.
User nodes can be connected to BS nodes that allow communications with the corresponding user group, providing the following three conditions are present:

\begin{enumerate}
\item When no users in the group have transmitted in the slot (or all the transmitted packets from the group in the slot have been retrieved), the user node has no edges.
\item If a single packet has been transmitted in the slot (or packets from multiple users have been transmitted and all except the packet of a single user have been retrieved), the user node and the BS nodes are connected with one edge.
\item The user node has two edges that extend to each connectable BS node when a collision has occurred. 

\end{enumerate}

\begin{figure}[t]
  \centering
  \includegraphics[width=0.25\hsize]{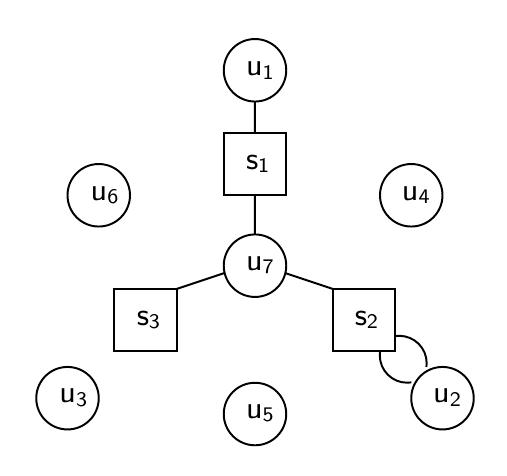}
  \vspace{-4ex}
  \caption{A walk graph of the frameless ALOHA with three cooperative BSs.}
  \vspace{-2ex}
  \label{walk_graph}
\end{figure}
Now, let us define three sets of user nodes, namely $\mathcal{U}_0$, $\mathcal{U}_1$, and $\mathcal{U}_2$, which include user nodes with no edges, a single edge, and two edges in the walk graph, respectively.
For example, the walk graph corresponding to the situation of Fig. \ref{edge_remain} is shown in Fig. \ref{walk_graph}, where $\mathcal{U}_0=\{\mathsf{u}_3,\mathsf{u}_4,\mathsf{u}_5,\mathsf{u}_6\}$, $\mathcal{U}_1=\{\mathsf{u}_1,\mathsf{u}_7\}$, and $\mathcal{U}_2=\{\mathsf{u}_2\}$.
The walk graph takes $3^I=3^{2^M-1}$ patterns, depending on the condition of each user node, and the probability of the instant walk graph can be obtained using degree distributions.
The probability that the user node has no edges with each connectable BS node is given by
\begin{equation}
  {\rm R}_i\triangleq R_i(1-x_i^{(l-1)}),\label{termR}
\end{equation}
and the probability that the user node has a single edge with each connectable BS node is 
\begin{equation}
  {\rm C}_i\triangleq \sum_{k=1}^{N_i}\binom{N_i}{k}p_i^k(1-p_i)^{N_i-k}kx(1-x)^{k-1}.\label{termC}
\end{equation}

Hence, the occurrence probability of the realization of walk graph $g$, such as $\mathrm{Pr}(g)$, is given by
\begin{equation}
  {\rm Pr}(g)=\prod_{i:\mathsf{u}_i\in \mathcal{U}_0}{\rm R}_i\prod_{j:\mathsf{u}_j\in \mathcal{U}_1}{\rm C}_j\prod_{k:\mathsf{u}_k\in \mathcal{U}_2}(1-{\rm R}_k-{\rm C}_k).
\end{equation}

When we focus on the user group $\mathsf{u}_u$, the probability that only the specified user will remain un-retrieved in the slot is given by
\begin{equation}
  {\rm r}_u\triangleq \rho_u(1-x_u^{(l-1)}).\label{termr}
\end{equation}

Therefore, given that only the specified user in $\mathsf{u}_u$ has transmitted, the probability of the walk graph $g$ is given by
\begin{equation}
  {\rm Pr}(g)={\rm r}_u\prod_{i:\mathsf{u}_i\in \mathcal{U}_0}{\rm R}_i\prod_{j:\mathsf{u}_j\in \mathcal{U}_1\setminus \{\mathsf{u}_u\}}{\rm C}_j\prod_{k:\mathsf{u}_k\in \mathcal{U}_2}(1-{\rm R}_k-{\rm C}_k).
\end{equation}

Note that the walk graph has $3^I$ patterns since there are $I$ user groups that accept three kinds of conditions depending on the number of edges.
Thus, the probability $P^{\rm (r0)}_i$ ($P^{\rm (r1)}_i$) can be obtained by calculating the sum of all the probabilities of walk graphs where the specified user can be retrieved from singleton slots (collided slots).
Let us define $\mathcal{R}^{\rm (r0)}_i$ as a set of walk graphs where the packet $\mathsf{u}_i$ can be retrieved from singleton slots, and $\mathcal{R}^{\rm (r1)}_i$ as a set of walk graphs where the packet can be retrieved from collided slots.
Then, $P^{\rm (r0)}_i$ and $P^{\rm (r1)}_i$ are obtained by $P^{\rm (r0)}_i=\sum_{g\in\mathcal{R}^{\rm (r0)}_i}{\rm Pr}(g)$ and $P^{\rm (r1)}_i=\sum_{g\in\mathcal{R}^{\rm (r1)}_i}{\rm Pr}(g)$.

Finally, $w_i^{(l)}$ is obtained by \eqref{w_i}, which can be rewritten as
\begin{align}
  w_i^{(l)}=1-\sum_{g\in\mathcal{R}_i}{\rm Pr}(g)\triangleq 1-P_i^{\rm (r)},\label{w_sum}
\end{align}
where $\mathcal{R}_i=\mathcal{R}^{\rm (r0)}_i\cup\mathcal{R}^{\rm (r1)}_i$.

Whether or not the instant walk graph belongs to $\mathcal{R}_i$ is determined by the use of an algorithm similar to SIC.
Since it is also a bipartite graph, SIC can be straightforwardly applied to the walk graph.
If the specified packet is retrieved, the instant graph is in $\mathcal{R}_i$.
Now, let us denote the degree of user node $i$ as $d_i$, and a universal set of walk graphs as $\mathcal{G}$.
Furthermore, we define $\mathcal{G}_{d_i=a}$ as a set of walk graphs that satisfy $d_i=a$.
Then, Algorithm I shows how to search all of the appropriate cases of $\mathsf{u}_i$ for the arbitrary number of $M$. 

Using the algorithm, the theoretical PLR performance and consequently the throughput performance can be derived for any number of BSs. This information is then used to obtain the multi-access diversity gain.
{\color{black} However, finding all the patterns where the specified packet can be retrieved is NP-hard, and the algorithm requires an exhaustive search over $3^I$ candidate walk graph patterns. Specifically, for $M=4$, the analysis requires evaluation over $10^7$ graphs, and the size increases to about $10^{14}$ when $M=5$. Hence, exact analyses involving cases of $M>4$ would be impossible due to their complexity. However, it is worth noting that we can utilize various search algorithms, $e.g.$ {\em backtracking}, to solve combinatorial search problems with low relatively complexity.}

\begin{algorithm}[h]
  \caption{Calculation of $w_i^{(l)}$}
  \begin{algorithmic}
    \STATE set $d_i=1$, $P^{\rm (r)}_i=0$.
    \FORALL{$g\in\mathcal{G}_{d_i=1}$}
    \STATE Carry out SIC on the walk graph $g$
    \IF{Succeeded}
    \STATE $P^{\rm (r)}_i+={\rm Pr}(g)$
    \ENDIF
    \ENDFOR
    \STATE $w_i^{(l)}=1-P^{\rm (r)}_i$
  \end{algorithmic}
\end{algorithm}
\vspace{-4ex}
\subsection{Approximated Analysis for General Case}\label{bound_analysis}
In order to alleviate the exponential complexity of calculating the exact throughput performance, we next derive the upper and lower throughput bounds for multi-BS cooperation.

\subsubsection{Upper Bound}
With a single BS, the highest throughput performance is 1.0, which can be achieved by TDMA\footnote{When a capture effect is available, multiple packets can be retrieved from a single time slot, resulting in a throughput performance higher than 1.0\cite{capture}.}.
Assuming a symmetric network with $M$ BSs where the numbers of users in every networks are same, the highest throughput performance is similarly $M$, which can be achieved when there exists no users in the overlapped coverage area.
This idea yields a simple upper bound of the throughput performance as follows.
Denote by $S_1$ the maximized throughput performance of frameless ALOHA with a single BS.
Specifically, it has been shown that frameless ALOHA asymptotically achieves a throughput of about $S_1=0.87$ \cite{Asilomar}.
Then, the throughput of frameless ALOHA with multiple BS cooperation is upper-bounded as
\begin{equation}
  S^{\rm c}\leq MS_1.\label{throughput_ub}
\end{equation}

Using the upper bound of the throughput, the upper bound of the multi-access diversity gain can be obtained by dividing \eqref{throughput_ub} by the throughput performance of a non-cooperative case obtained using \eqref{pe_nc}.
\subsubsection{Lower Bound}
In order to obtain the lower throughput bound, we derive the {\em upper} PLR bound for frameless ALOHA with cooperative multiple BSs by assuming a toy model as follows.
When focusing on the packet $\mathsf{u}_i$, only the retrieved packets of the users in $\mathsf{u}_i$ are shared among the BSs, not the packets of users in the other groups.
This is equivalent to ignoring the probability $P_i^{\rm (r1)}$ in \eqref{w_i}.
Hence, from the viewpoint of throughput, this toy model is obviously inferior to the actual network model where multiple BS cooperation is available.
In the following paragraphs, we explain how the lower bound of the throughput is derived by the upper bound of the PLR.
The upper bound of the PLR is obtained by bounding $P^{\rm (r)}_i$ from the {\em lower} side, where we use the lower bound for a union probability, as proposed in \cite{Kounias}, to reduce the resulting computational complexity.
Consider a union probability of $n$ random variables, $i.e.$, the probability that at least one of $n$ events, namely $A_i,i\in[1,n]$, has occurred, which is represented by ${\rm Pr}(\bigcup_{i=1}^nA_i)$.
Using a $1\times n$ vector $\mathbf{p}=({\rm Pr}(A_1),\ldots,{\rm Pr}(A_n))$ and an $n\times n$ square matrix $\mathbf{Q}=\{{\rm Pr}(A_i\cap A_j)\}$, the union probability is lower-bounded as
\begin{equation}
  {\rm Pr}(\bigcup_{i=1}^nA_i) \geq \mathbf{p}\mathbf{Q}^{-1}\mathbf{p}^{\rm t},\label{lb}
\end{equation}
where $\mathbf{Q}^{-1}$ is an inverse matrix of $\mathbf{Q}$, and $\rm t$ denotes transposition \cite{Kounias}.

We were motivated to employ the lower bound of \eqref{lb} not only because it is elegant, but also because it is more accurate than the well-known Bonferroni lower bound.
Consider applying the bound to the calculation of $P^{\rm (r)}_i$.
The probability $P^{\rm (r)}_i$ can be rewritten as the union probability, specifically
\begin{equation}
  P^{\rm (r)}_i={\rm Pr}(\bigcup_{j:\mathsf{s}_j\in\mathcal{S}(\mathsf{u}_i)}\mathsf{u}_i\text{ is retrieved at }\mathsf{s}_j).
\end{equation}

Then, we can apply the lower bound to the calculation of $P^{\rm (r)}_i$.
Based on the toy model introduced given above, the probability of the packet $\mathsf{u}_i$ being retrieved at $\mathsf{s}_j$ is given by
\begin{align}
  P_i^{\rm (r)}(\mathsf{s}_j)= {\rm r}_i\prod_{k:\mathsf{u}_k\in\mathcal{U}(\mathsf{s}_j)\setminus\{\mathsf{u}_i\}}{\rm R}_k\label{p_ret1}
\end{align}
where $P_i^{\rm (r)}(\mathsf{s}_j)$ denotes the probability that the packet $\mathsf{u}_i$ has been retrieved at $\mathsf{s}_j$, and $\mathcal{U}(\mathsf{s}_j)$ is a set of user groups that has been set up so that all group users can communicate with $\mathsf{s}_j$.

The probability corresponds to cases where the packet $\mathsf{u}_i$ is retrieved at $\mathbf{s}_j$ since the packet is received without any collisions in the actual network model.
It is obvious that the bound ignores cases where the specified packet $\mathsf{u}_i$ is retrieved after the packet has collided.
Moreover, based on the toy model, the joint probability corresponding to cases where the packet $\mathsf{u}_i$ is retrieved at both $\mathsf{s}_{j_1}$ and $\mathsf{s}_{j_2}$ is given by
\begin{equation}
  P_i^{\rm (r)}(\mathsf{s}_{j_1},\mathsf{s}_{j_2})= {\rm r}_i\prod_{k:\mathsf{u}_k\in\mathcal{U}(\mathsf{s}_{j_1})\cup\mathcal{U}(\mathsf{s}_{j_2})\setminus\{\mathsf{u}_i\}}{\rm R}_k\label{p_ret2}
\end{equation}
where $P_i^{\rm (r)}(\mathsf{s}_{j_1},\mathsf{s}_{j_2})$ denotes the probability that the packet $\mathsf{u}_i$ is retrieved at both $\mathsf{s}_{j_1}$ and $\mathsf{s}_{j_2}$.
By substituting \eqref{p_ret1} and \eqref{p_ret2} into \eqref{lb}, where elements of $\mathbf{p}$ and diagonal elements of $\mathbf{Q}$ are given by \eqref{p_ret1}, and non-diagonal elements of $\mathbf{Q}$ are given by \eqref{p_ret2}, we can obtain the lower bound for $P_i^{\rm (r)}$.
Note that the size of $\mathbf{Q}$ is $|\mathcal{S}(\mathsf{u}_i)|\times|\mathcal{S}(\mathsf{u}_i)|$, and none of the diagonal elements of $\mathbf{Q}$ are zero as long as $G_i>0$.
Using \eqref{p_ret1} and \eqref{p_ret2}, the upper PLR bound can be derived by lower-bounding $P_i^{\rm (r)}$.

Thanks to the simplification of \eqref{p_ret1} and \eqref{p_ret2}, an exhaustive search over $3^{2^M-1}$ candidates in the exact analysis is replaced with only, at most, $M(M+1)/2$ terms\footnote{The number of terms varies depending on how many BSs the specified user group can communicate with. For $\mathsf{u}_i$, the number of terms is $|\mathcal{S}(\mathsf{u}_i)|(|\mathcal{S}(\mathsf{u}_i)|+1)/2$.}.
On the other hand, the simplification may loosen the upper bound of the PLR.
As mentioned above, in order to obtain the exact PLR performance, all of the patterns should be taken into account.
Hence, if we want to make the bound tighter, we need to consider more terms, which requires a larger computational cost, and it is hard to formulate a tight bound at a low computational cost.
However, the fact remains that the upper bound of the PLR ($i.e.$, the lower bound of the throughput) reveals the guaranteed performance gain, and hence, the lower throughput bound is valuable.

\section{Numerical Examples}\label{numerical_example}
In this section, we examine frameless ALOHA with multiple BS cooperation and give some numerical examples.
Specifically, the optimization problem on target degrees, which maximizes the average throughput, is introduced.
The optimization problem is based on the previously derived theoretical throughput expressions.
Furthermore, the performance improvement via multiple BS cooperation, namely {\em multi-access diversity gain}, was evaluated, and it was revealed that the gain increases almost linearly as the number of BSs increases.
Moreover, frameless ALOHA using the optimized target degrees is compared with a state-of-the-art random-access scheme that also uses multiple BS cooperation, with the results showing that the optimized frameless ALOHA scheme significantly outperforms the conventional multiple BS cooperation scheme thanks to the frameless structure and the exact analysis of PLR.

\subsection{Target Degree Optimization}
The throughput at the $T$-th time slot $S(T)$ has been defined as the fraction of retrieved information packets and elapsed time slots, which is shown in \eqref{definition_throughput}.
The equation of \eqref{definition_throughput} can be rewritten as $S(T)=\sum_i N_i(1-p_{{\rm e},i}(T))/T$.
The average throughput performance of the original (single-BS) frameless ALOHA peaks as the number of time slots increases, and it has been shown that the actual throughput performance converges to the theoretical performance as the number of users increases. Additionally, the average throughput performance converges to the peak throughput value when there is a sufficiently large number of users \cite{Peter_lett,Asilomar}.
 Hence, in order to maximize the average throughput, the optimization problem is formulated to find the target degree vector $\mathbf{G}_{\rm opt}=\{G_{1,\mathrm{opt}},\ldots,G_{I,\mathrm{opt}}\}$ which maximizes the peak throughput, that is
\begin{align}
  \max_{\bf G} &\quad\sup_T S(T)\label{optimization}\\
  \mathrm{s.t.} &\quad1-p_{\rm e}(T^*)>\alpha,
\end{align}
where $T^*=\mathrm{arg\ sup}_TS(T)$.

Using the theoretical analysis of PLR, target degrees can be optimized for arbitrary networks with an arbitrary number of BSs.
Consider the target degree optimization for networks with $I=2^M-1$ user groups with $M\leq 4$.
The threshold is set to $\alpha=0.8$ and the frame terminates when 80\% of all the packets have been successfully retrieved.
Without the loss of generality, it is constrained that $G_i=G_j$ for all $i$ and $j$ such that $|\mathcal{S}(\mathsf{u}_i)|=|\mathcal{S}(\mathsf{u}_j)|$.
The optimization problem is a multi-modal problem with multiple target degrees to be optimized. In this paper, the differential evolution \cite{DE1}, which is regularly used to optimize the degree distribution of graph-based codes such as LDPC codes \cite{density-evolution}, is employed to solve the problem.
In the optimization example, $300$ candidates are used, $0.2$ is used as the mutant factor, and the update on candidates (generating mutants) are iterated 30 times.
Details on the differential evolution algorithm can be found in \cite{DE1}.
In the subsections below, we first study optimization using several BS numbers while considering a symmetric network.
Then, in order to show the relationship between the number of users in each network and the optimal target degrees, optimization for an asymmetric network is considered.

\subsubsection{Optimization for Symmetric Networks}

\begin{table}[t]
  \begin{center}
    \caption{The optimal target degrees for $M\leq 4$.}
    \begin{tabular}{|c||c|c|c|c||c|c|} \hline
      & $|\mathcal{S}(\mathsf{u}_i)|=1$ & $|\mathcal{S}(\mathsf{u}_i)|=2$ & $|\mathcal{S}(\mathsf{u}_i)|=3$ & $|\mathcal{S}(\mathsf{u}_i)|=4$ & Theoretical peak throughput & Simulated average throughput \\ \hline \hline
      $M=1$ & 3.10 &  -   &  -   &  -   & 0.874 & 0.867 \\\hline
      $M=2$ & 1.81 & 1.68 &  -   &  -   & 1.676 & 1.673 \\\hline
      $M=3$ & 1.11 & 0.94 & 0.78 &  -   & 2.366 & 2.363 \\ \hline
      $M=4$ & 0.69 & 0.52 & 0.46 & 0.46 & 2.940 & 2.936 \\\hline
    \end{tabular}
    \label{opt_results}
  \end{center}
  \vspace{-7ex}
\end{table}

First, let us consider symmetric networks where $N_i=10^4$ for all $i$.
Table \ref{opt_results} shows the optimal target degrees and corresponding throughput performance for different BS numbers. Note that achievable throughput with $M$ BSs is $M$, at most. In our computer simulation, we presumed that $N_i=10^4$ for each user group and confirmed that the theoretical peak throughput performance coincides with the actual average throughput performance of the computer simulation. The reason why the simulated performance values degrade slightly from theoretical performance values is that the theoretical analysis assumes the number of graph nodes to be infinite, so that the graph becomes typical.
When multiple BSs exist in the network, target degrees of user groups in overlapped areas become smaller than that of the isolated user group.
These results agree with the conclusion in \cite{Corson}, where it is suggested that users in overlapped areas should not transmit too much.
Moreover, these results contradict the results of \cite{Jakovetic}, where it is suggested that all the users should transmit with an equal probability in the presence of  an SIC with multiple cooperating BSs.
Throughput comparison with \cite{Jakovetic} will discussed in Section \ref{comparison_spatio_temporal}.

\subsubsection{Optimization for Asymmetric Networks}

It may be of interest to note how the optimal target degrees and corresponding throughput performance vary when the number of users in each group differs.
To clarify the discussion so far, let us consider networks with $M=2$ BSs and consequently $I=2^M-1=3$ user groups. Specifically, $\mathsf{u}_1$, $\mathsf{u}_2$, and $\mathsf{u}_3$.
Groups-1 and 2 are able to communicate with BS-1 and 2 respectively, and $\mathsf{u}_3$ can communicate with both BSs.
Moreover, for the sake of simplicity, it is assumed that $N_1=N_2$ and $G_1=G_2$ so that we only need to mention $N_1$, not both $N_1$ and $N_2$.
We optimized target degrees for several networks and listed the results in Table \ref{AsymOpt}.
\begin{table}[t]
  \begin{center}
    \caption{The optimal target degrees for several networks.}
    \begin{tabular}{|c||c|c||c|c||c|c|} \hline
      Network type & $N_1$ & $N_3$ & $G_1$ & $G_3$ & Theoretical peak throughput & Simulated average throughput\\\hline
      (a) & 0 & 10,000 & - & 3.098 & 0.874 & 0.867 \\\hline
      (b) & 100 & 10,000 & 1.388 & 3.094 & 0.893 & 0.890 \\\hline
      (c) & 1,000 & 10,000 & 1.621 & 3.063 & 1.064 & 1.060 \\\hline
      (d) & 10,000 & 10,000 & 1.812 & 1.680 & 1.676 & 1.673 \\\hline
      (e) & 10,000 & 1,000 & 3.051 & 1.869 & 1.836 & 1.829 \\\hline
      (f) & 10,000 & 100 & 3.096 & 0.302 & 1.758 & 1.746 \\\hline
      (g) & 10,000 & 0 & 3.098 & - & 1.748 & 1.736 \\\hline
    \end{tabular}
    \label{AsymOpt}
  \end{center}
  \vspace{-7ex}
\end{table}
We confirmed that theoretical peak throughput performances agree with simulated average throughput performances.
In the following, the results are discussed in terms of two kinds of situations: $N_3>N_1$ and $N_3<N_1$.

When $N_3>N_1$, where the number of users in the overlapped area is larger than that of users in isolated area, the peak throughput performance degrades as $N_1$ decreases.
This is natural because the network approaches  single BS network, where all the users can communicate with common BSs.
Note that since network (a) is identical to a single BS network, the optimal target degree and the peak throughput are also identical to the original frameless ALOHA.
The optimal target degree of $\mathsf{u}_3$ approaches the optimal value for single BS frameless ALOHA. $i.e.$, 3.098, as $N_1$ decreases.
When it comes to $\mathsf{u}_1$, the optimal target degree $G_1$ is decreased in order to bring it into balance with the increase of $G_3$. This is necessary because if the users in $\mathsf{u}_1$ transmit frequently, the channel would become saturated and the SIC process would be stacked.
Hence, in order to a achieve high throughput performance, the optimal $G_1$ must be decreased as the optimal $G_3$ in $N_3>N_1$ is increased,
As in the discussion above, when $N_3<N_1$, the optimal target degree of $\mathsf{u}_1$ approaches the optimal target degree of the original frameless ALOHA as $N_3$ decreases.
In the most extreme case, $i.e.$, network (g), the optimal target degree is identical to the original frameless ALOHA, and the peak throughput is twice that of the original frameless ALOHA.
Interestingly, the peak throughput values of networks (e) and (f) are both higher than that of (g).
This is because packets transmitted from users in overlapped areas may be retrieved by other BSs while the average traffic load at each BS is the same as that for a single BS frameless ALOHA with an optimized target degree.
An observation we can obtain here is that the number of users affects the throughput improvement via multiple BS cooperation.
It is especially noteworthy that the existence of some users in overlapped areas increases rather than degrades throughput performance.

\subsection{Comparison with Some Simple Schemes}
We further show how much performance improvement is achieved by the proposed frameless ALOHA compared to non-cooperative frameless ALOHA schemes.
For comparison purposes, two simple transmission schemes employing frameless ALOHA are considered while supposing that $M=2$.
Without loss of generality, let us consider again symmetric networks.

The simplest case is {\em perfect separation}, where each BS has its own frame.
First, the users in $\mathsf{u}_1$ and a portion of the users in $\mathsf{u}_3$, namely $\mathsf{u}_{3,1}$ with $N_{3,1}$ users, transmit packets to $\mathsf{s}_1$ with frameless ALOHA. Then, after the frame of $\mathsf{s}_1$, the users in $\mathsf{u}_2$ and the remaining users in $\mathsf{u}_3$, namely $\mathsf{u}_{3,2}$ with $N_{3,2}$ users such that $N_{3,1}+N_{3,2}=N_3$, transmit to $\mathsf{s}_2$.
Perfect separation is equivalent to a frequency division model, where each BS uses its own frequency band.

Another case is {\em simultaneous transmission}, where all of the users transmit in the same frame while BSs retrieve packets without multiple BS cooperation.
These two candidates use the optimal target degree $G=3.098$, which is designed for $M=1$, since BSs do not cooperate.

\begin{figure}[t]
  \centering
  \includegraphics[width=0.48\hsize]{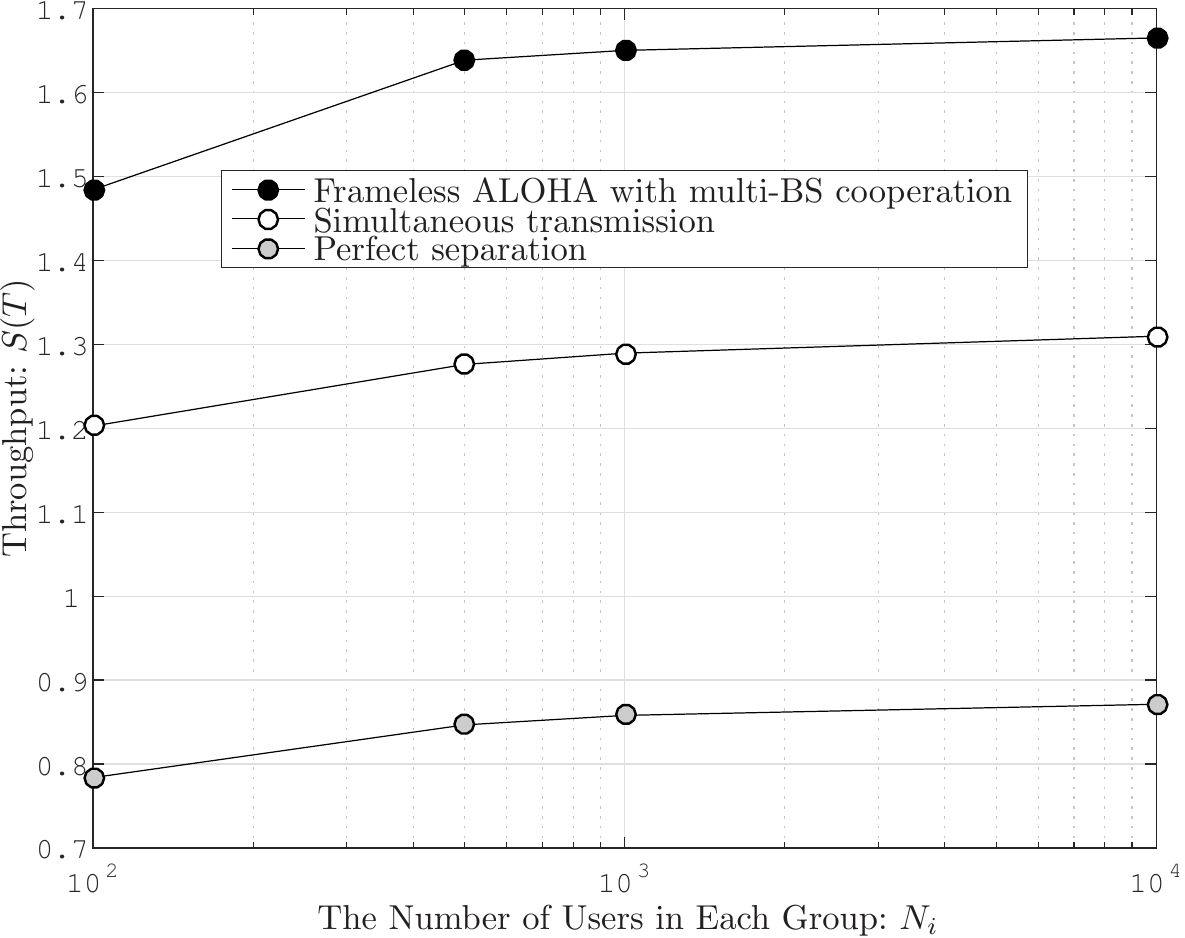}
  \caption{Comparison of throughput performance of frameless ALOHA with multiple BS cooperation, perfect separation, and simultaneous transmission.}
  \label{throughput_diversity}
  \vspace{-4ex}
\end{figure}

Figure \ref{throughput_diversity} shows the throughput performance of each scheme.
Simultaneous transmission performs better than perfect separation because its features allow user in $\mathsf{u}_1$ and $\mathsf{u}_2$ to transmit simultaneously without interfering with each other and without any performance degradation. 
However, when combined with multiple BS cooperation and optimized target degrees, frameless ALOHA exhibits a throughput performance that is clearly higher than simultaneous transmission.
An important observation to keep in mind is that the use of multiple cooperating BSs significantly improves the throughput performance due to the existence of \emph{multi-access diversity gain}, which exploits the overlapped coverage areas.
Furthermore, in the presence of SIC and multiple BS cooperation, coverage overlapping improves the throughput performance because of multi-access diversity gain, which is in contradiction to the results in the classic multiple BS random access literature \cite{Corson, David}.
However, it should also be noted that this contradiction arises because the literature had not  yet considered interference cancellation, and the previously given advice that users in overlapped areas should be separated remains partially sound because excessive overlapping decreases throughput performance.
Therefore, it can be concluded that, when multiple BS cooperation is available, multiple cells should be merged to facilitate maximum performance, rather than keeping each cell separate.
We also examined multi-access diversity gain from the viewpoint of theoretical analysis.
Recall that the main difference between cooperative and non-cooperative analysis is the calculation of $w_i^{(l)}$.
From \eqref{w_i}, the probability $w_i^{(l)}$ consists of two components: $P_i^{\rm (r0)}$ and $P_i^{\rm (r1)}$.
In particular, the latter term corresponds to the probability that packets can be retrieved thanks to multiple BS cooperation, which cancels colliding packets.
This is the exact outcome desired from multiple BS cooperation, and it is predicted that $P_i^{\rm (r1)}$ {\em contributes significantly to multi-access diversity gain}.
\begin{figure}[t]
  \centering
  \includegraphics[width=0.48\hsize]{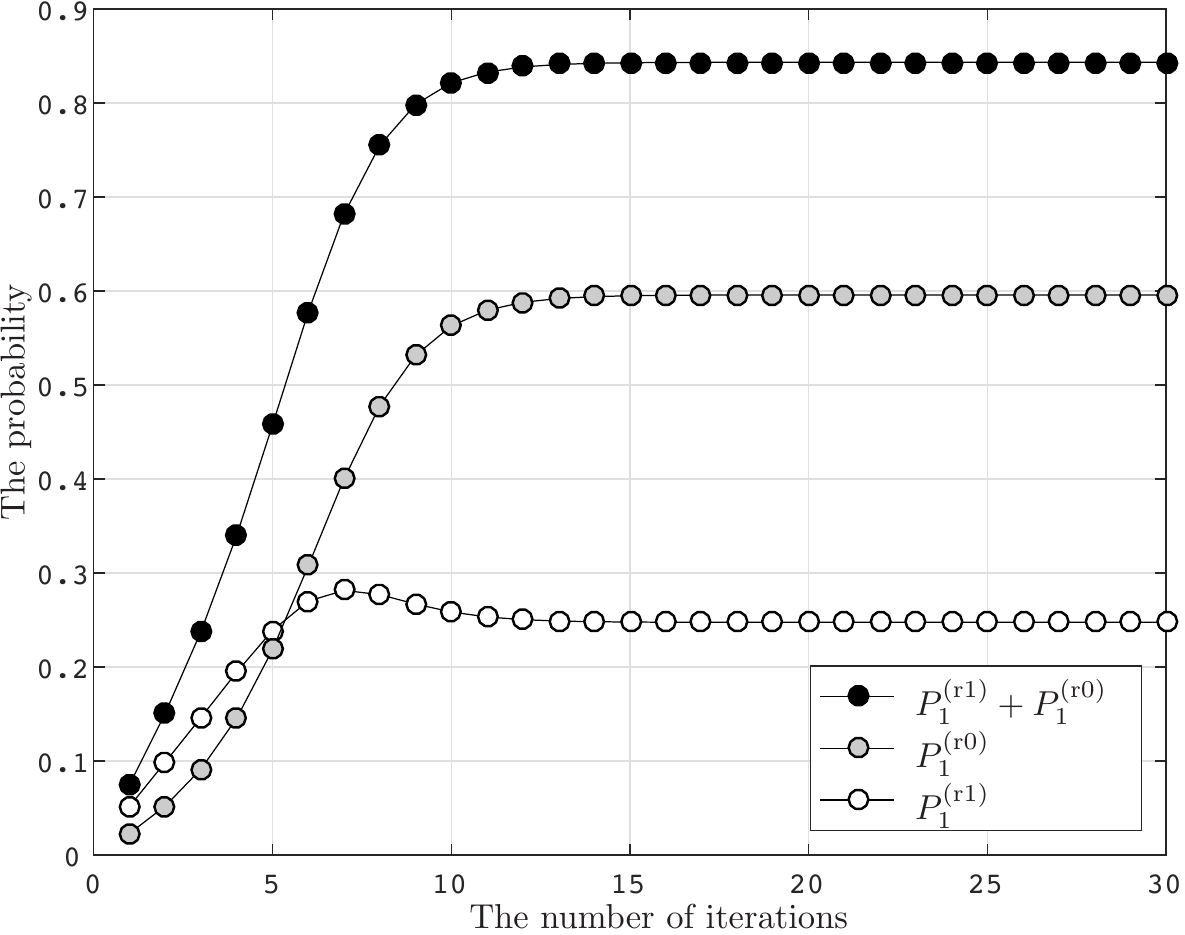}
  \caption{Evolution of $P_i^{\rm (r0)}$ and $P_i^{\rm (r1)}$ in the iterative calculation.}
  \label{evolution_of_p}
  \vspace{-4ex}
\end{figure}
Figure \ref{evolution_of_p} shows the evolution of $P_1^{\rm (r0)}$ and $P_1^{\rm (r1)}$ in the iterative calculation of $w_i^{(l)}$ and $x_i^{(l)}$, supposing that $M=3$.
Note that the resulting PLR decreases as $P_i^{\rm (r0)}$ and $P_i^{\rm (r1)}$ increase, which is obvious from \eqref{w_i}.
Interestingly, in contrast to the prediction, $P_i^{\rm (r0)}$ plays a more important role than $P_i^{\rm (r1)}$.
Another important result of multiple BS cooperation is that packet sharing among BSs cancels more packets than in the non-cooperative case, thereby resulting in more singleton slots.
When multiple BS cooperation is not available, each BS locally performs the SIC, as previously elucidated.
In contrast, with cooperation, the SIC is performed {\em jointly} among BSs, thereby leading to the efficient retrieval of packets from overlapped areas.
The idea is similar to the diversity techniques of a physical layer.
One insight we can report here is that the gain is available as long as multiple cooperative BSs exist.
  
\subsection{Evaluation of Multi-Access Diversity Gain}

It has been confirmed that multiple BS cooperation enhances throughput performance more than separated-BS systems.
Using \eqref{eq:gain}, we will now examine how the multi-access diversity gain varies with $M$.
Note that $S^{\rm nc}$ corresponds to the throughput of {\em simultaneous transmission} in the previous subsection, since simultaneous transmission achieves higher throughput than perfect separation.

\begin{figure}[t]
  \centering
  \includegraphics[width=0.48\hsize]{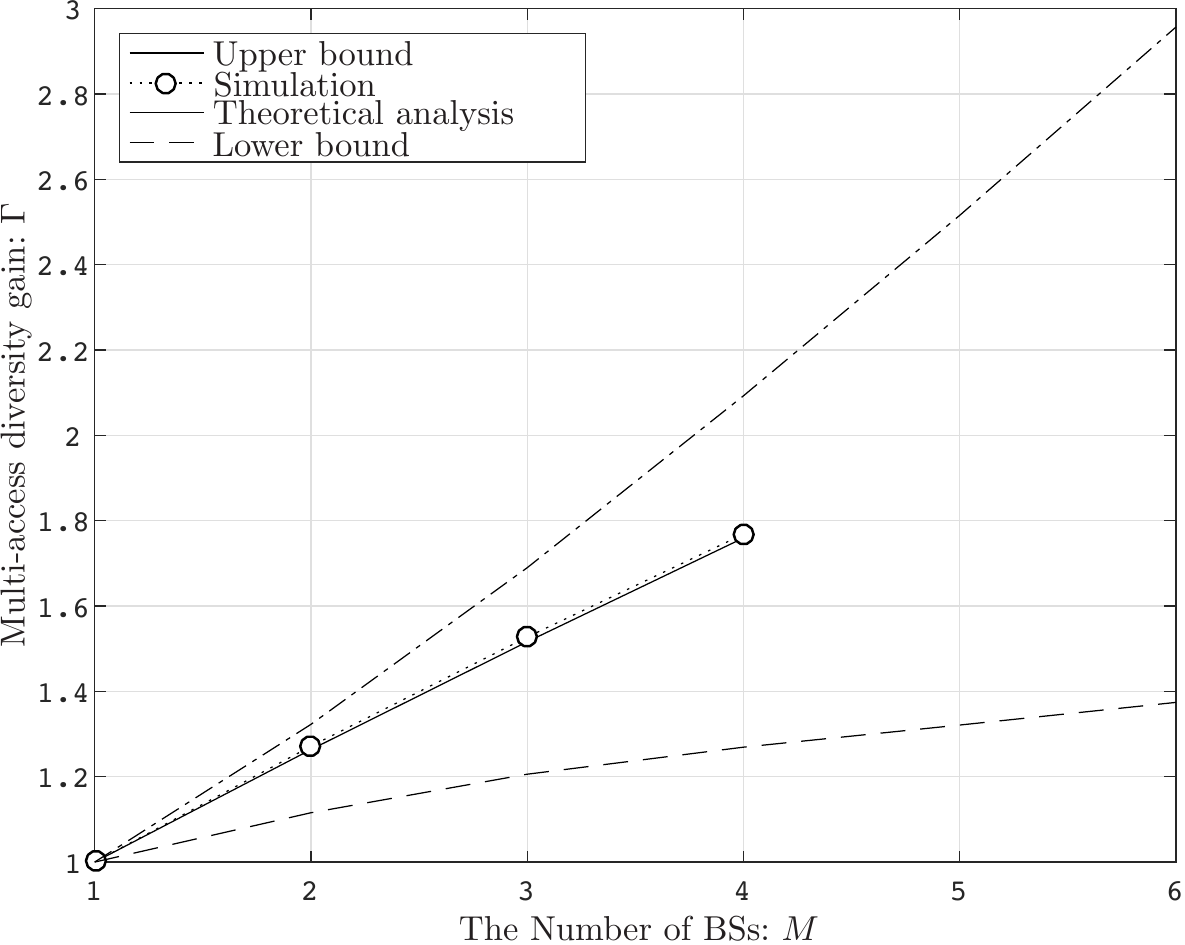}
  \caption{Multi-access diversity gain $\Gamma$ as a function of the number of BSs $M$.}
  \label{multiaccess_diversity_gain}
  \vspace{-4ex}
\end{figure}
In Fig. \ref{multiaccess_diversity_gain}, the multi-access diversity gain $\Gamma$ is depicted as a function of the number of BSs $M$, as well as the lower and upper bounds.
The lower and upper bounds are given in Section \ref{bound_analysis}.
Note that target degrees used to obtain the lower bound are optimized point-by-point using the PLR upper bound.
We confirmed that our proposed theoretical analysis shows good agreement with the results of computer simulations.
Due to computational costs, exact theoretical analysis and computer simulation results have only been obtained for $M\leq 4$.
Although the upper bound is simple, we can obtain a meaningful insight that the achievable gain is at most linear.
This is important, since the achievable limitation of the multiple BS cooperation is explicitly shown and we do not have to discover parameters such that the throughput is higher than $MS_1$.
In other words, the upper bound can be seen as a {\em capacity} of the frameless ALOHA with multiple BS cooperation; the bound is an extreme goal which the system can achieve. 
Furthermore, it is worth noting that the increasing rate of the gain is maximized when the target degrees are optimized.

We also determined that the lower bound given by \eqref{lb}--\eqref{p_ret2} is not tight.
Although the bound is based on the lower bound of union probability, we can also make lower bounds by calculating \eqref{w_sum} using an arbitrary subset $\mathcal{Q}_i\subset\mathcal{R}_i$ instead of $\mathcal{R}_i$.
At that point, the accuracy of the bound can be controlled by the size of $\mathcal{Q}_i$.
The bound becomes identical to the exact analysis when $\mathcal{Q}_i=\mathcal{R}_i$, and the accuracy degrades as the size of $\mathcal{Q}_i$ decreases.
However, as the number of BSs increases, producing accurate bounds requires enormous computational complexity.
To this end, an affordable option is to use the matrix-based bound as the lower-bound of throughput performance for frameless ALOHA with multiple BS cooperation.
Although the lower bound is not tight, the bound is remarkable since the bound {\em strictly exceeds} 1.0; we can {\em always increase} the throughput performance by multiple BS cooperation, and it is guaranteed that the throughput performance never degrades by the cooperation.
This is a motivational insight, since we can obtain higher throughput gain as we deploy larger number of BSs. 

It may be interesting to study how the multi-access diversity gain varies when the number of users in each group differs.
For comparison purposes, we focus on three networks picked from Table \ref{AsymOpt}: (c) $N_1=N_2=10^3,N_3=10^4$, (d) $N_1=N_2=N_3=10^{4}$, and (e) $N_1=N_2=10^4,N_3=10^3$.
Recall that the gain of the symmetric network (d) is $\Gamma=1.26$.
In network-(e), where the number of users in the overlapped area is {\em smaller} than the number of isolated users, $S^{\rm c}=1.84$, $S^{\rm nc}=1.66$, and $\Gamma=1.11$.
The gain is smaller than that of the symmetrical network, meaning that the effect of multiple BS cooperation is slightly less than that of the symmetrical network.
This is natural because the number of users that can be retrieved via multi-BS cooperation is smaller than can be retrieved by the symmetrical network.
On the other hand, in network-(c), where the number of users in the overlapped area is {\em larger} than the number of isolated users, $S^{\rm c}=1.06$, $S^{\rm nc}=0.97$, and $\Gamma=1.09$.
The gain also decreases in comparison to the case of the symmetrical network because most of the users in the network belong to the overlapped area, which means that the network performance approaches that of a single BS case.
\subsection{Effect of System Parameters}
In this section, the effect of system parameters, $i.e.$, the number of users and the threshold, on the optimal target degree and multi-access diversity gain is discussed.
\subsubsection{Effect of the Number of Users}
The effect of the asymmetric network has been discussed previously.
Note that the optimal target degree is affected by the {\em ratio} of the number of users in each group, but not the specific number of users.
That is to say, the resulting optimal target degrees will be the same for two cases where $N_i=10^4$ for all $i$ and $N_i=10^5$ for all $i$, as long as the number of users in each group is sufficiently large.
If the number of users is small, then the assumption of density evolution that guarantees that the degree distribution of the graph is typical does not hold, leading to incorrect results\footnote{Finite length analysis for frameless ALOHA can be found in \cite{Lazaro}.}.

\subsubsection{Effect of Threshold $\alpha$}
Recall that $\alpha$ is the packet retrieval ratio required to finish the transmission frame.
If $\alpha$ is too low, then the results would be equivalent to considering only the achieved throughput performance, but not the resulting PLR.
Such optimization obviously results in {\em unfair} target degrees, which means that some user groups cannot deliver their packets to the BSs.
On the other hand, optimization with too high an $\alpha$ yields poor throughput performance because frameless ALOHA uses probabilistic transmission, which leads to an error floor.
That is, if we set the threshold at an extremely high value such as $\alpha=1.0-10^{-5}$, obviously the frame length $T$ must be large enough to achieve the required PLR, thereby resulting in decreased throughput performance.

Although the design of a practically optimal $\alpha$ was investigated in \cite{Peter_trans}, our primary interest is to theoretically design and analyze the system. Thus, detailed discussions about parameters, $e.g.$, at what value should $\alpha$ be practically set in order to achieve a high throughput, are beyond the scope of this paper.
\subsection{Comparison with State-of-the-Art}\label{comparison_spatio_temporal}
A framed ALOHA protocol employing multiple BS cooperation and SIC, named {\it spatio-temporal cooperation}, was proposed in \cite{Jakovetic}, and to the best of our knowledge, that scheme remains the state-of-the-art random access structure involving multiple BS cooperation.
In this section, we have shown that our proposed frameless ALOHA with multiple BS cooperation achieves a higher throughput performance than spatio-temporal cooperation.
In spatio-temporal cooperation, each user selects a temporal degree $s$ according to a {\it degree distribution} $\Lambda=(\Lambda_1, \cdots, \Lambda_{s_{\rm max}})$ at the beginning of the frame, where $\Lambda_s$ is the probability that the degree $s$ is selected, and the maximum degree is denoted by $s_{\rm max}$.

\begin{figure}[t]
  \begin{center}
    \includegraphics[width=0.45\hsize]{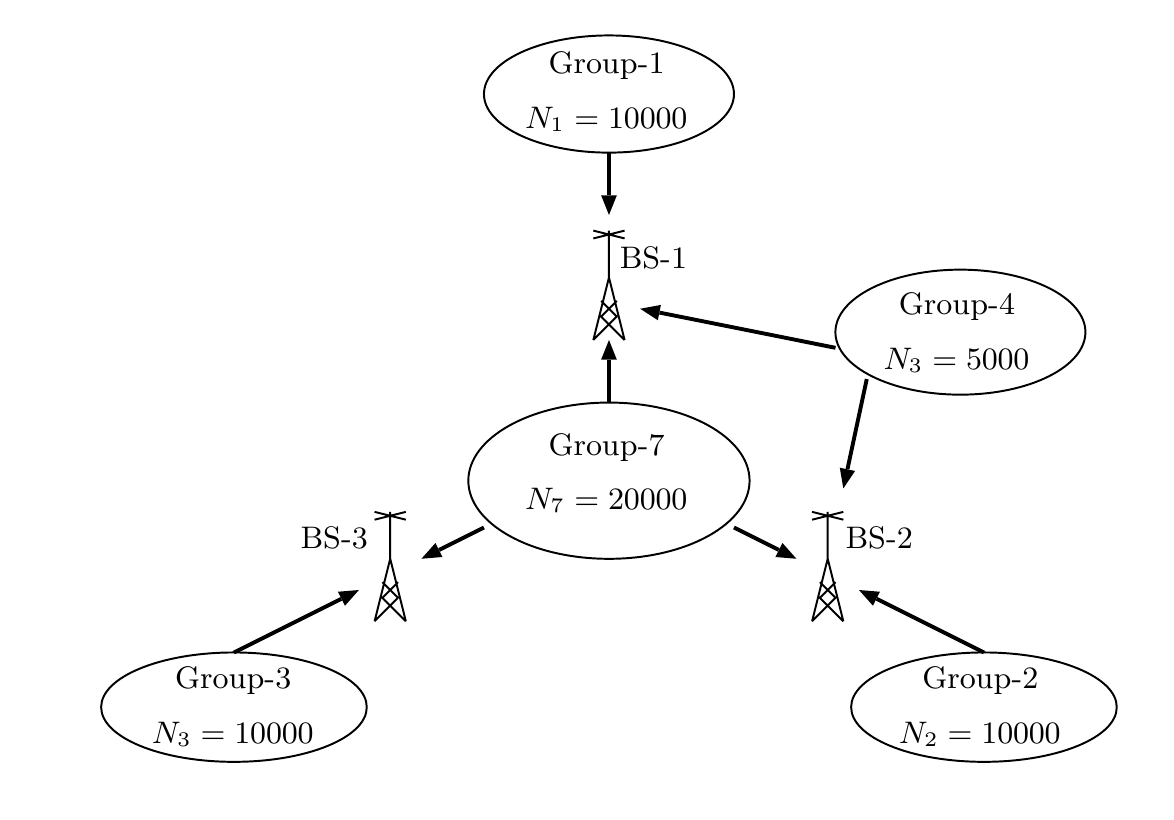}
    \vspace{-6ex}
    \caption{The network model for $\delta=2$.}
    \label{model_compare}
  \end{center}
\vspace{-4ex}  
\end{figure}

The frame length is fixed $a\ priori$ so that the user with the degree $s$ transmits $s$ times during the frame.
Each user's transmission is heard by multiple BSs, and the average {\em spatial degree} $\delta$ is defined as the average number of BSs that can receive a packet from the user.
Upon transmission, the BSs attempt to retrieve the transmitted packets using SIC, and successfully retrieved packets are shared among all the BSs.
In \cite{Jakovetic}, the degree distribution is optimized for some $\delta$.

In order to compare our proposed frameless ALOHA with spatio-temporal cooperation, let us consider the network of $\delta=2$, as shown in Fig. \ref{model_compare}.
For $\delta=2$, the optimal degree distribution given in \cite{Jakovetic} is $\Lambda^*_2=1$, which means that all the users transmit two times during the frame.
The optimal target degrees of frameless ALOHA for the network can be obtained by the optimization problem given in \eqref{optimization}, in which we need to optimize target degrees $G_1,G_2,G_3,G_4$, and $G_7$.
Since $\mathsf{u}_1$ and $\mathsf{u}_2$ are symmetric with $\mathsf{s}_1$ and $\mathsf{s}_2$, $i.e.$, both groups are connected to a single BS that the users in $\mathsf{u}_4$ and $\mathsf{u}_7$ attempt to use simultaneously, it is assumed that $G_1=G_2$.

The obtained optimal target degrees are $(G_1,G_3,G_4,G_7)=(1.42,1.30,0.47,2.33)$.
For comparison purposes, the normalized throughput performance versus the normalized load is evaluated for both schemes.
The normalized load $\overline{G}$ is given by dividing the number of users in the network by the number of time slots and BSs, that is
\begin{equation}
  \overline{G}\triangleq \frac{N}{MT}.
\end{equation}

\begin{figure}[t]
  \centering
  \includegraphics[width=0.48\hsize]{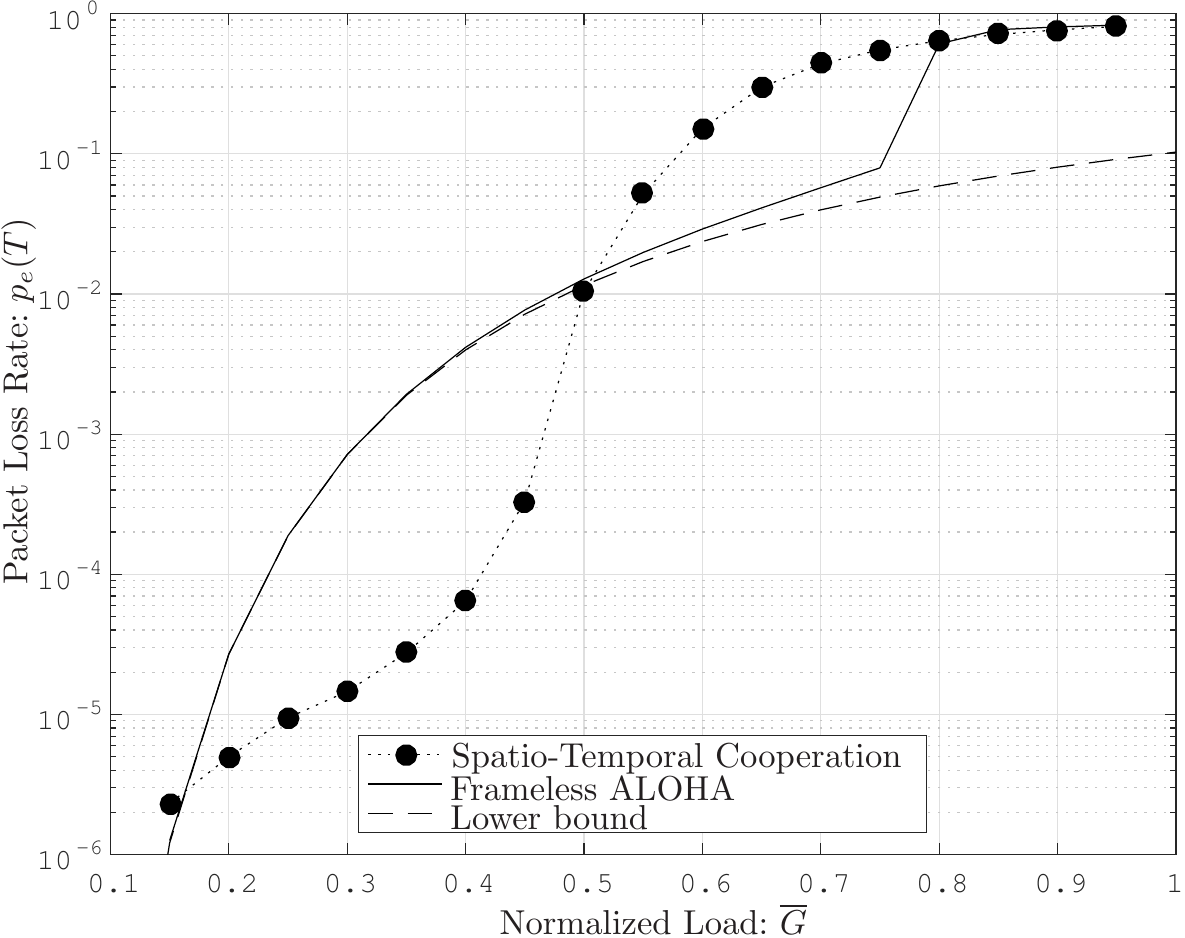}
  \caption{The average PLR performance of frameless ALOHA with multiple BS cooperation and spatio-temporal cooperation.}
  \label{norm_plr}
  \vspace{-4ex}
\end{figure}

Normalized throughput $\overline{S}(\overline{G})$ is given by dividing throughput by the number of BSs, that is
\begin{equation}
  \overline{S}(\overline{G})\triangleq \frac{\sum_i N_i(1-p_{{\rm e},i}(T))}{MT}.
\end{equation}

\begin{figure}[t]
  \centering
  \includegraphics[width=0.48\hsize]{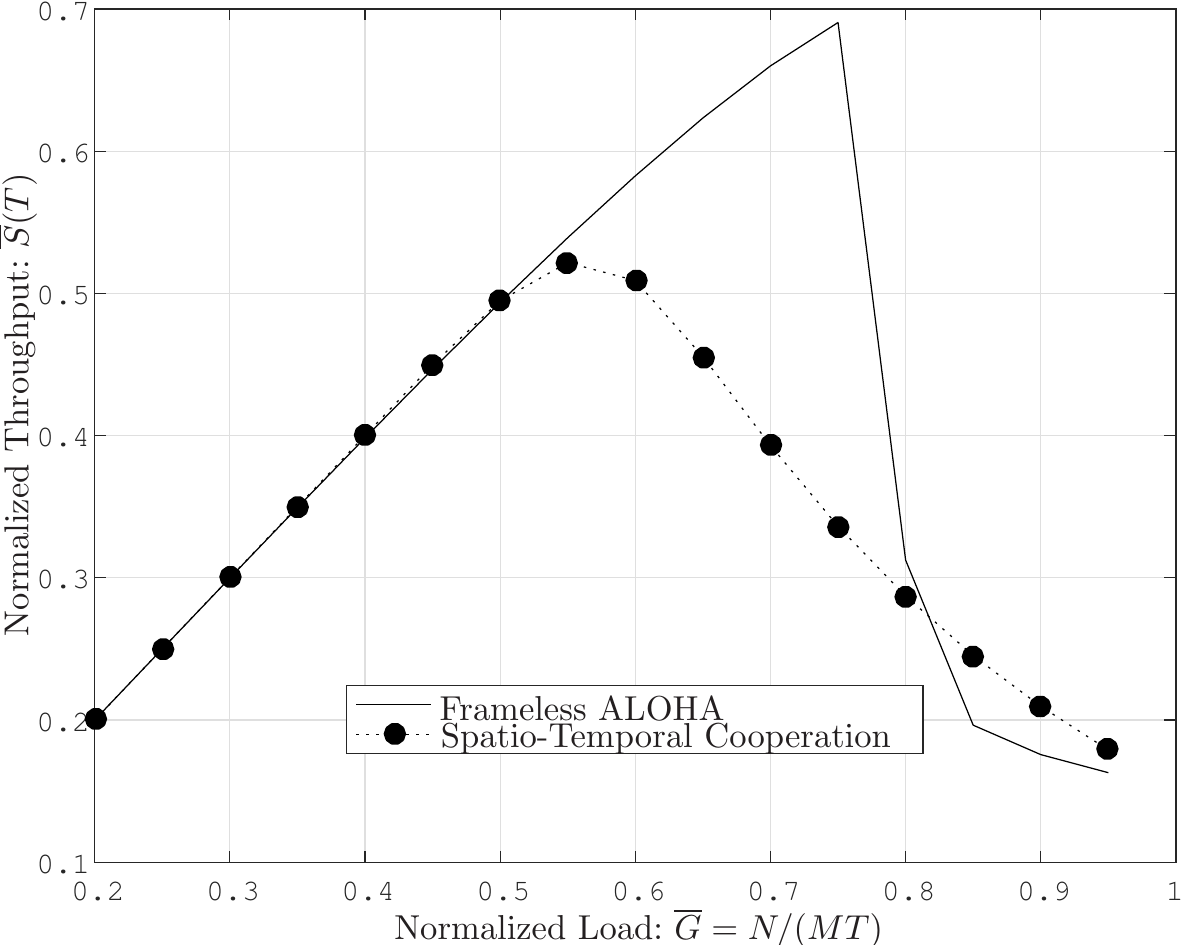}
  \vspace{-3ex}
  \caption{The normalized throughput performance of frameless ALOHA with multiple BS cooperation and spatio-temporal cooperation for the network model shown in Fig. \ref{model_compare}.}
  \label{norm_throughput}
  \vspace{-4ex}
\end{figure}

Figures \ref{norm_plr} and \ref{norm_throughput} show, respectively, the normalized throughput performance and the average PLR performance of frameless ALOHA with multiple BS cooperation and spatio-temporal cooperation obtained via computer simulations.
The PLR performance of frameless ALOHA is lower-bounded by the probability that the user never transmits in the frame, given by
\begin{equation}
  p_{{\rm e}_{\rm LB}}(T)=\sum_{i=1}^I\frac{N_i}{N}(1-p_i)^T.\label{fa_lb}
\end{equation}

It is worth noting that the PLR performance suddenly drops at approximately $\overline{G}=0.8$.
The steep fall is called a {\em waterfall region}, and is a representative characteristic of a belief propagation (BP) decoder \cite{Urbanke}, and the SIC process can be interpreted as a BP decoder for binary erasure channels (BEC).
Frameless ALOHA with multiple BS cooperation using the aforementioned optimized target degrees beats spatio-temporal cooperation from the viewpoint of both throughput and PLR in a practical area $0.5 \leq \overline{G}\leq 0.8$.
Our observations show that our proposed scheme is capable of bearing a heavier load, and thus can achieve a higher throughput performance, $i.e.$, while the performance of spatio-temporal cooperation degrades after the point of $G=0.55$, frameless ALOHA with the optimal target degrees achieves a high throughput until the load reaches the point of $G=0.75$.
{\color{black}Due to the approximation in the analysis of spatio-temporal cooperation, the conventional scheme can only assign a single degree distribution to all the users. In contrast, frameless ALOHA with our proposed analysis can employ different transmission probabilities for each user group, which means that our proposed scheme outperforms the conventional scheme.}

For the area with a smaller load that is $G<0.5$, frameless ALOHA has worse PLR performance than spatio-temporal cooperation. However, the throughput performance of frameless ALOHA is almost the same as that of spatio-temporal cooperation because the PLR gap in the region is lower than $10^{-2}$.
In the heavier load area, $i.e.$, $G>0.8$, the proposed frameless ALOHA performs worse than spatio-temporal cooperation.
This is because the saturated channel makes the offered load too heavy to carry out random access schemes.
The comparison here, however, has fixed the number of time slots for both frameless ALOHA and conventional schemes, even though frameless ALOHA adaptively determines the number of time slots.
In other words, frameless ALOHA always adjusts the channel load to the optimal point where the peak throughput is achieved by its frameless structure.
{\color{black}Specifically, frameless ALOHA adaptively determines its frame length so as to retrieve sufficiently large number of users by terminating the frame when the number of retrieved users exceeds the {\em a priori} given threshold. This means that the protocol finishes when peak throughput is achieved. Moreover, thanks to the exact throughput analysis utilized in the optimization process, our proposed frameless ALOHA has a peak throughput performance which is higher than the average throughput performance of the conventional scheme.}
Thus, we can say that our frameless ALOHA scheme outperforms a state-of-the-art random access structure for use with multiple BS cooperation.
\section{Conclusion}\label{conclusion}
In this paper, we examined frameless ALOHA with multiple BS cooperation and showed how much performance improvement is achievable by defining the multi-access diversity gain.
An exact theoretical analysis of the throughput performance was given, along with a simple lower bound.
Theoretical analysis was used to optimize target degrees so that the achievable throughput could be maximized.
Numerical examples have shown that the multi-access diversity gain monotonically increases as the number of BSs increases.
Moreover, we confirmed that our proposed frameless ALOHA scheme with optimal target degrees outperforms the conventional coded MAC scheme in the sense of normalized throughput performance.

\appendix[Formulas of $w_i^{(l)}$]\label{appendix}
The calculation of $w_i^{(l)}$ used for $M=3$ taking into consideration all the cases where colliding edges can be immediately canceled at other BSs is available.
Using \eqref{termR} and \eqref{termC}, and let $\bar{{\rm R}}_i\triangleq 1-{\rm R}_i$, $w_i^{(l)}$ is given by
\begin{align}
  w_1^{(l)}=&1-{\rm r}_1\left({\rm R}_4{\rm R}_6{\rm R}_7+{\rm C}_4{\rm R}_2{\rm R}_5{\rm R}_6{\rm R}_7+{\rm C}_4{\rm C}_5{\rm R}_2{\rm R}_3{\rm R}_6{\rm R}_7+{\rm C}_6{\rm R}_3{\rm R}_4{\rm R}_5{\rm R}_7+{\rm C}_6{\rm C}_5{\rm R}_2{\rm R}_3{\rm R}_4{\rm R}_7\right.\notag\\
    +&\left.{\rm C}_4{\rm C}_6{\rm R}_2{\rm R}_3{\rm R}_5{\rm R}_7+{\rm C}_7\left({\rm R}_4{\rm R}_5{\rm R}_6\left(1-\bar{{\rm R}}_2\bar{{\rm R}}_3\right)+{\rm C}_4{\rm R}_2{\rm R}_3{\rm R}_5{\rm R}_6+{\rm C}_6{\rm R}_2{\rm R}_3{\rm R}_4{\rm R}_5\right) \right),\label{w1_3BSs}
\end{align}
\begin{align}
  w_4^{(l)}=&1-{\rm r}_4({\rm R}_7({\rm R}_5{\rm R}_6(1-\bar{{\rm R}}_1\bar{{\rm R}}_2)+(1-{\rm R}_5-{\rm C}_5){\rm R}_1{\rm R}_6+(1-{\rm R}_6-{\rm C}_6){\rm R}_2{\rm R}_5\notag\\
    +&{\rm C}_5{\rm R}_6({\rm R}_1+\bar{{\rm R}}_1{\rm R}_2{\rm R}_3)+{\rm C}_6{\rm R}_5({\rm R}_2+\bar{{\rm R}}_2{\rm R}_1{\rm R}_3))+{\rm C}_7{\rm R}_3{\rm R}_5{\rm R}_6(1-\bar{{\rm R}}_1\bar{{\rm R}}_2)),\label{w4_3BSs}
\end{align}
\begin{align}
  w_7^{(l)}=&1-{\rm r}_7({\rm R}_4{\rm R}_5{\rm R}_6(1-\bar{{\rm R}}_1\bar{{\rm R}}_2\bar{{\rm R}}_3)+{\rm R}_4{\rm R}_5\bar{{\rm R}}_6{\rm R}_2+{\rm R}_4\bar{{\rm R}}_5{\rm R}_6{\rm R}_1+\bar{{\rm R}}_4{\rm R}_5{\rm R}_6{\rm R}_3).\label{w7_3BSs}
\end{align}

\bibliography{IEEEabrv,/Users/shun/Dropbox/Paper/bib_file}
\end{document}